\documentclass[twocolumn,aps,showpacs,prl,superscriptaddress]{revtex4-1}
\usepackage{amsmath}
\usepackage{amssymb}
\usepackage{graphicx}
\usepackage{bbold}
\usepackage{times}
\usepackage{pdfpages}
\usepackage[normalem]{ulem}

\usepackage[colorlinks=true,allcolors=blue,bookmarks=false,pdfusetitle]{hyperref}
\usepackage{url}

\makeatletter

\usepackage{color}

\newcommand{\sgn}{\mbox{sgn}}

\makeatother

\begin{document}
\title{Collective many-body bounce in the breathing-mode oscillations of a 
Tonks-Girardeau gas}
\author{Y.~Y.~Atas}
\affiliation{University of Queensland, School of Mathematics and Physics,
Brisbane, Queensland 4072, Australia}

\author{I. Bouchoule}
\affiliation{Laboratoire Charles Fabry, Institut d'Optique, CNRS, Univesit\'e Paris
Sud 11, 2 Avenue Augustin Fresnel, F-91127 Palaiseau Cedex, France}

\author{D.~M.~Gangardt}
\affiliation{School of Physics and Astronomy, University of Birmingham, Edgbaston,
Birmingham, B15 2TT, UK}

\author{K.~V.~Kheruntsyan}
\affiliation{University of Queensland, School of Mathematics and Physics,
Brisbane, Queensland 4072, Australia}

\date{\today}

\begin{abstract}
We analyse the breathing-mode oscillations of a harmonically quenched Tonks-Giradeau (TG) gas using an exact finite-temperature dynamical theory.  
We predict a striking collective manifestation of 
  impenetrability---a collective many-body bounce effect. The effect,
    while being invisible in the evolution of the \emph{in situ} density profile of the gas,  can be
 revealed through a nontrivial periodic narrowing of its momentum distribution, taking place at twice the rate of the fundamental breathing-mode frequency.  We identify physical regimes for observing the many-body bounce and construct the respective nonequilibrium phase diagram as a function of the quench strength and the initial temperature of the gas. We also develop a finite-temperature hydrodynamic theory of the TG gas, wherein the many-body bounce is explained by an increased thermodynamic pressure during the isentropic compression cycle, which acts as a potential barrier for the particles to bounce off. 
  \end{abstract}


  \maketitle
  
Collective dynamics in many-body systems emerge as a result of interparticle interactions. Such dynamics can be characterised by a coherent or correlated behaviour of the constituents, which cannot be predicted form the single-particle or noninteracting picture. Collective dynamics can therefore serve as an important probe of the underlying interactions and is at the heart of a variety of nonequilibrium phenomena in many-body physics, including the archetypical examples of superfluidity and superconductivity. 
Among physical systems of current theoretical and experimental interest for understanding nonequilibrium many-body dynamics are ultracold quantum gases \cite{Kinoshita2006,hofferberth2007,trotzky2012,gring2012,fang2014quench,Bloch-Dalibard-Zwerger,CazalillaRigolNJP2010,polkovnikov2011,Cazalilla2011,lamacraft2012}, which offer a versatile platform for realising minimally complex but highly controllable models of many-body theory.

In quantum gases, the simplest manifestations of collective dynamics relate to the frequencies of monopole (breathing-mode) and multipole oscillations in harmonic trapping potentials \cite{Pitaevskii-Stringary-book,Stringari:1996,Pitaevskii:1998,JILA1996,MIT1996,JILA1997,MIT1998,Moritz2003,Naagerl2009,fang2014quench,Hinds2016,JILA2016}. These frequencies, depending on trap configurations, can vary significantly from those of ideal (noninteracting) gases. 
For example, in a weakly interacting 1D Bose gas at sufficiently low temperatures, the breathing-mode oscillations of the \emph{in situ} density occur at frequency $\omega_{B}\simeq \sqrt{3}\omega$ (where $\omega$ is the frequency of the trap) \cite{Moritz2003,Naagerl2009,fang2014quench,Minguzzi-hydro-2001,Menotti:2002,Hu2014,*Hu2015,Choi:2015,Astrakharchik-Zvonarev-2015,Stringari2015,*Stringari2016,Bouchoule_qbec:2016}, whereas in an ideal Bose gas the breathing-mode frequency is $\omega_{B}=2\omega$. An even more dramatic, quantitative departure from the ideal gas behaviour was recently observed in the dynamics of the momentum distribution of a weakly interacting 1D quasicondensate \cite{fang2014quench,Bouchoule_qbec:2016}: for sufficiently low temperatures, the momentum distribution was oscillating at frequency $2\omega_B$, i.e., at twice the rate of the fundamental breathing-mode frequency of the \emph{in situ} density profile, $\omega_{B}\simeq \sqrt{3}\omega$. Furthermore, at intermediate temperatures the oscillations could be decomposed as a weighted superposition of just two harmonics, one oscillating at $2\omega_B$ and the other at $\omega_B$.

In a strongly interacting 1D Bose gas, on the other hand, the breathing-mode oscillations are predicted to display the so-called re-entrant behaviour \cite{Menotti:2002,Naagerl2009,Choi:2015,Astrakharchik-Zvonarev-2015}, wherein the frequency of the \emph{in situ} density oscillations returns to the value characteristic of the ideal Bose gas. This implies that a single-particle behaviour is seemingly recovered, even though the system is strongly interacting. This behaviour can be understood by the fact that in the extreme limit of infinitely strong interactions or the Tonks-Girardeau (TG) regime \cite{Girardeau1960,*Girardeau_Bose_Fermi,Girardeau2000,*yukalov2005fermi}, the 1D Bose gas can be mapped to a system of noninteracting fermions, which---just as the ideal Bose gas---oscillates at the fundamental breathing-mode frequency of $\omega_{B}=2\omega$, both in real and momentum spaces.

In this work, we show that the collective behaviour in the breathing oscillations of the TG gas can nevertheless be revealed via the dynamics of its momentum distribution. Here, the collective dynamics manifest itself as a many-body bounce effect, which is absent in the ideal Fermi gas and is characterized by periodic narrowing of the momentum distribution that occurs, as in the weakly interacting case, at twice the rate of oscillations of the \emph{in situ} density. In contrast to the weakly interacting case \cite{fang2014quench,Bouchoule_qbec:2016}, however, the periodic narrowing at the inner turning points occurs on relatively short time scales so that the oscillations cannot be generally represented as a superposition of just two harmonics, except in the regime of extremely small oscillation amplitudes. Despite this difference, our findings imply that the many-body bounce is a universal emergent property of the breathing-mode dynamics of 1D Bose gases, generic to both weak and strong interactions. 

Our analysis is based on the exact finite-temperature dynamical theory of the TG gas developed recently \cite{TonksMethods} using the Fredholm determinant approach. We apply this theory to construct a nonequilibrium phase diagram for the many-bounce effect, parametrized in terms of the dimensionless initial temperature and the quench strength.
Our findings are further supported by a finite-temperature hydrodynamic theory of 1D Bose gases \cite{Bouchoule_qbec:2016}, which we apply here to the TG gas dynamics in a harmonic trap. Interestingly, the hydrodynamic scaling solutions emerge here as soon as one invokes the local density approximation on the exact many-body solutions, without any further assumptions (such as, e.g., fast thermalization rates).

\begin{figure*}[th]
\includegraphics[width=15.5cm]{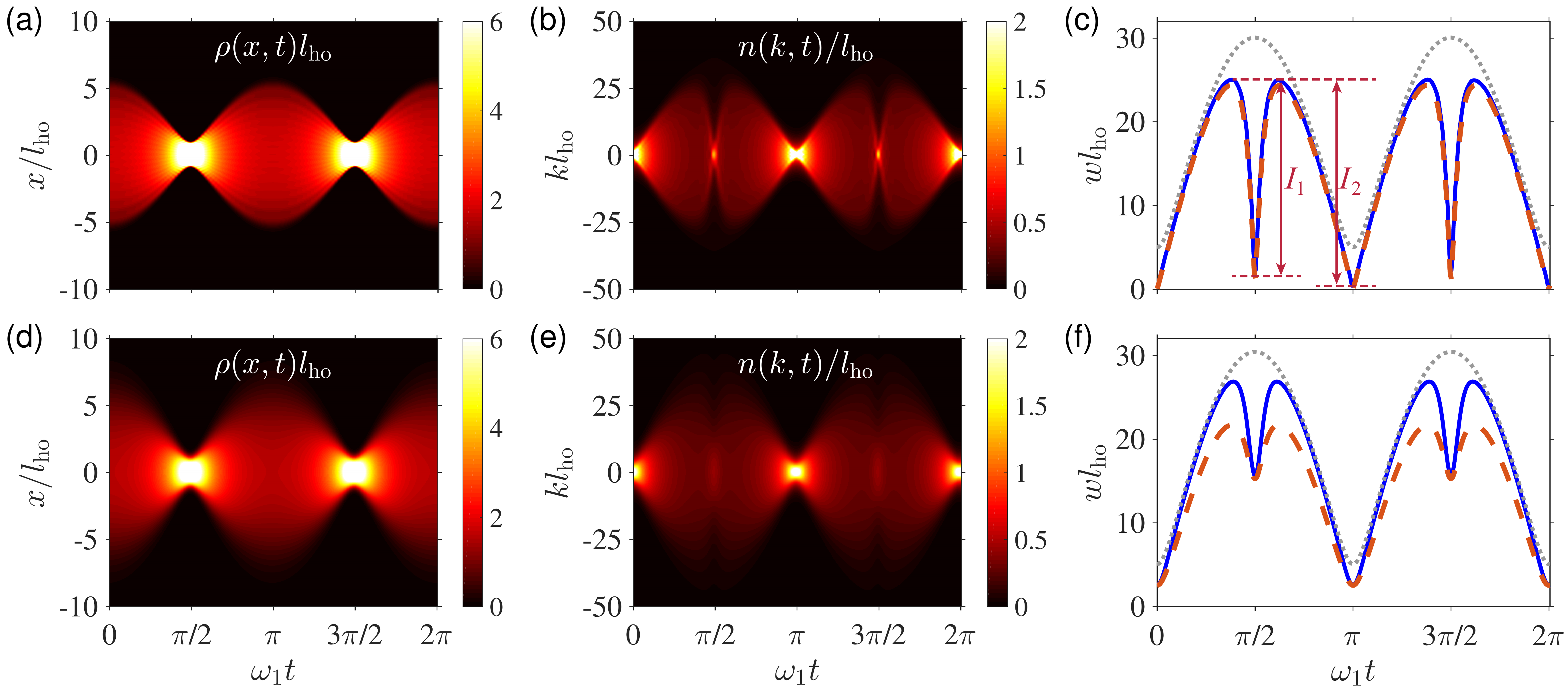}
 \caption{(Color online) Breathing-mode dynamics of the TG gas following a confinement 
     quench.  (a) Real-space density distribution, $\rho(x,t)l_{\mathrm{ho}}$; (b) momentum distribution, $n(k,t)/l_{\text{ho}}$; and (c) the width (HWHM) of the momentum distribution, $w(t)$ (with $wl_{\text{ho}}$ being dimensionless), as functions of the dimensionless time $\omega_1t$, for $N\!=\!16$ particles, quench strength $\epsilon\!\simeq\!-0.9722 $ ($\omega_1\!=\!6\omega_0$), and dimensionless initial temperature $\theta_0 \!\equiv \!k_{B}T_0/N\hbar\omega_{0}\!=\!0.01$. 
  (d)--(f) Same as before, but at higher temperature, $\theta_0\!=\!0.5$.
     We use the harmonic oscillator length $l_{\mathrm{ho}}\!=\!(\hbar/m\omega_0)^{1/2}$ as the lengthscale. In (c) and (f) the solid (blue) lines are from the 
    exact calculations, the dotted (grey) lines are the momentum width of an ideal Fermi gas shown for comparison, and the dashed (orange) lines are from the 
    hydrodynamic theory (see text).  In (c), $I_1$ and $I_2$ indicate the depths of the local and global minima in the width of the momentum distribution (see text).}
\label{fig:big}
\end{figure*}

We start by recalling that the TG gas corresponds to a system of $N$ impenetrable (hard-core) bosons of mass $m$ \cite{Girardeau1960,*Girardeau_Bose_Fermi}, which we assume are confined in a time-dependent harmonic trap $V(x,t) = m\omega(t)^2 x^2/2$, where $\omega(t)$ is the trap frequency. 
The problem of its evolution can be solved exactly \cite{TonksMethods} by employing the Bose-Fermi mapping \cite{Girardeau1960,*Girardeau_Bose_Fermi,Girardeau2000,*yukalov2005fermi,vignolo2013universal}, which reduces the interacting many-body problem to a single-particle basis of a noninteracting Fermi gas. If the trapping potential remains harmonic at all times, the reduced one-body density matrix of the TG gas $\rho(x,y;t)$ can be obtained from the initial one $\rho_0(x,y)\equiv \rho(x,y;0)$ by a scaling transformation
\cite{PerelomovBook,GangardtMinguzziExact},
\begin{equation}
\rho(x,y;t)=\frac{1}{\lambda}
\rho_0\left(x/\lambda,y/\lambda\right) e^{i m\dot\lambda (x^2-y^2)/2\hbar \lambda},
\label{eq.ss}
\end{equation}
where the scaling
parameter $\lambda(t)$ is to be found from the 
ordinary differential equation (ODE)
$\ddot{\lambda}=-\omega(t)^{2}\lambda+\omega_{0}^{2}/\lambda^{3}$, with the
initial conditions
 $\lambda(0)\!=\!1$, and $\dot{\lambda}(0)\!=\!0$.

In order to observe the quantum many-body bounce effect, we invoke the breathing-mode
oscillations using a confinement quench, in which at $t\!=\!0$ 
the trapping frequency $\omega(t)$ is instantaneously changed  
from the pre-quench value
$\omega_0$ to a new value $\omega_1$. In this case, the ODE 
for $\lambda(t)$ acquires the form of the Ermakov-Pinney equation,
$\ddot{\lambda}=-\omega_1^{2}\lambda+\omega_{0}^{2}/\lambda^{3}$, with the
solution
\begin{eqnarray}
  \label{eq:solution_scaling}
  \lambda(t) = \sqrt{1 +\epsilon \sin^2(\omega_1 t)},
\end{eqnarray}
where $\epsilon\equiv \omega^{2}_{0}/\omega^{2}_{1}-1$ is the quench strength.

The scaling solution (\ref{eq.ss}) simplifies the analysis enormously as the one-body density matrix needs to be calculated only once, at time $t=0$. 
To calculate $\rho_0(x,y)$, we use the following 
 exact and computationally practical expression, found recently \cite{TonksMethods} using the Fredholm determinant approach and valid at arbitrary temperatures:
 \begin{equation}
\rho_0(x,y)={\textstyle \sum_{i,j=0}^{\infty}}\sqrt{f_{i}}
\phi_{i}(x)Q_{ij}(x,y)\sqrt{f_{j}}\phi_{j}^{\ast}(y). 
\label{Finite_temp_densitymat_Tonks}
\end{equation}
Here, $f_j=[e^{(E_j-\mu)/k_BT_0}+1]^{-1}$ is the Fermi-Dirac distribution
function for the single-particle orbital occupancy, described by the wavefunction $\phi_j(x)$ and energy $E_j=\hbar\omega_0(j+1/2)$, $T_0$ is the initial equilibrium temperature, and $\mu$ is the chemical potential. In addition,
$Q_{ij}$ are the  
matrix elements of the operator 
$\mathbf{Q}(x,y)=(\mathbf{P}^{-1})^{\mathsf{T}}\mathrm{det}\;\!\mathbf{P}$, with
\begin{equation}
 P_{ij}(x,y)\! =\! \delta_{ij}- 2\;\!\sgn(y-x)\sqrt{f_{i}f_{j}} \!\int_{x}^{y}
 \!\!\!dx^{\prime}\phi_{i}(x^{\prime})\phi_{j}^{\ast}(x^\prime).
  \label{matrix_element}
\end{equation}

Combining Eqs.~(\ref{Finite_temp_densitymat_Tonks})--(\ref{matrix_element}) with~(\ref{eq.ss})--(\ref{eq:solution_scaling}) allows
one to calculate important observables, such as real-space
density $\rho(x,t) \!=\! \rho(x,x;t)$ and momentum distribution
$n(k,t)\!=\!\int dx\, dy \, e^{-ik(x-y)}\rho(x,y;t)$ of the TG gas.  
Evolution of these quantities after a strong quench ($\omega_1\!=\!6\omega_0$, $\epsilon\!\simeq\!-0.9722$) is shown in
Fig.~\ref{fig:big} for $N\!=\!16$ particles and two different initial temperatures $T_0$.
The dynamics of the real-space density, given initially by
$\rho_0(x)\!=\!\rho(x,0)$, consist of self-similar broadening/narrowing (breathing)
cycles, $\rho(x,t)\! =\! \rho_0(x/\lambda)/\lambda$,
 always occurring at the
fundamental breathing-mode frequency of $\omega_B\!=\!2\omega_1$, independently of
$T_0$.  In contrast, the evolution of the momentum distribution $n(k,t)$
is not self-similar and displays a more complicated structure that depends on
$T_0$ 
and $\epsilon$.

\begin{figure}[tbp]
 \includegraphics[width=7.8cm]{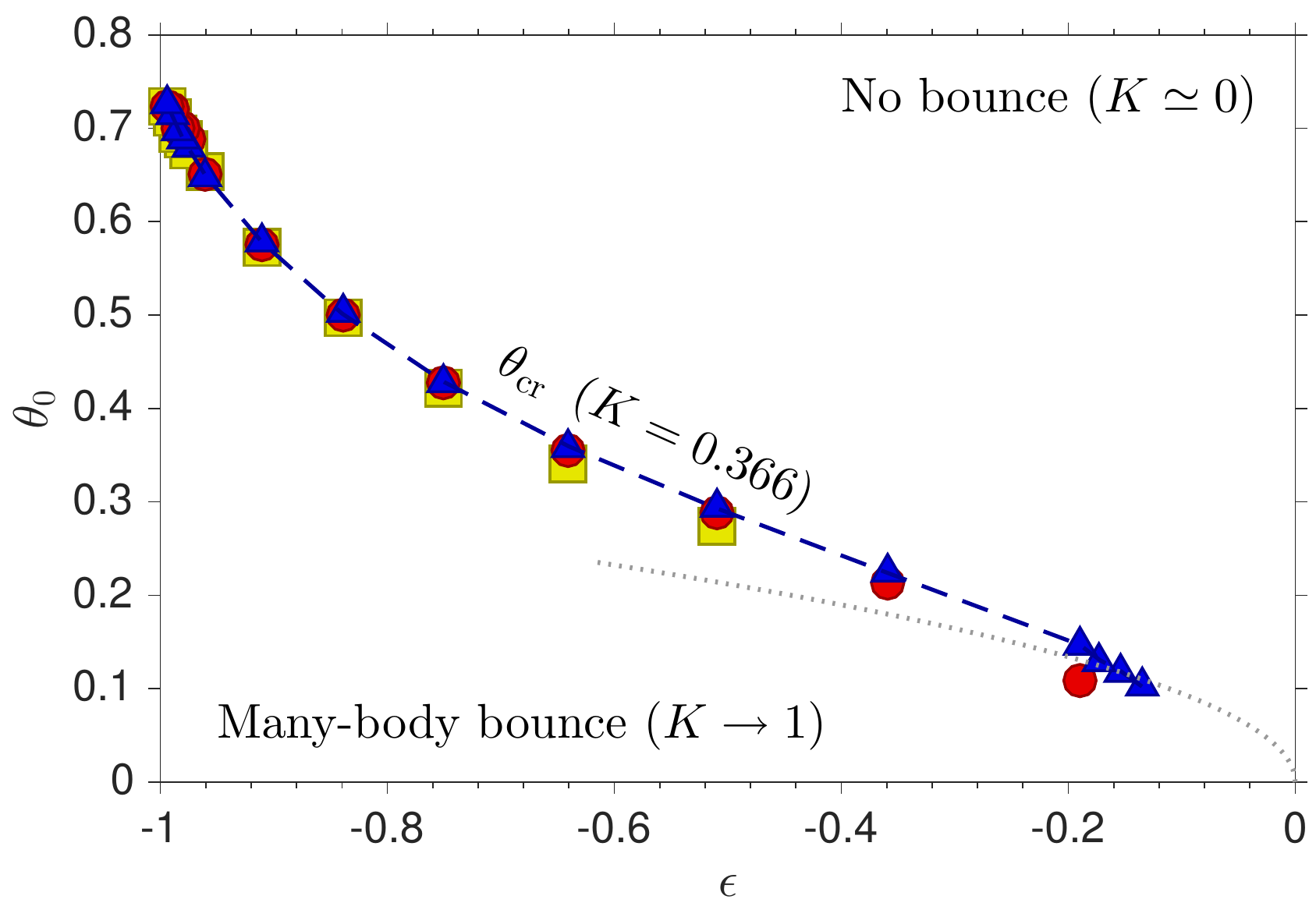}
     \caption{ (Color online) Crossover phase diagram for the phenomenon of quantum many-body bounce. The data points from the exact theory show the locations of the crossover temperature 
         $\theta_{\text{cr}}$, for $N\!=\!8$ (squares), $12$ (circles), and $16$ (triangles). Due to finite size effects, 
the value of $K$, for any fixed $N$, always stays smaller than the crossover value of $K\!=\!0.366$ 
until a sufficiently strong quench (the strength itself being dependant on $N$) is applied, hence the absence of 
data points for $\theta_{\mathrm{cr}}$, for, e.g., $N\!=\!8$ in the region $-0.5\! \lesssim \!\epsilon \! < \!0$.
  The  dashed line connecting the triangles is drawn to guide the eye.
The dotted line shows the analytic prediction of $\theta_{\text{cr}}\!\simeq \!0.3 \sqrt{|\epsilon|}$ for $|\epsilon|\!\ll\! 1$.} 
\label{Fig_Crossover_condition}
\end{figure}


The many-body bounce effect manifests itself as a visible narrowing of the
momentum distribution at time instances corresponding to $\omega_1t\!=\!\pi l$
($l\!=\!1,2,...$), when the gas is maximally compressed and the impenetrable bosons
slow down and reverse their momenta near the bottom of the trap (for an illustration of the breathing-mode 
dynamics for $N\!=\!2$, see~\cite{sup-Bounce}).
These instances of narrowing,
which we refer to as \emph{inner} turning points, occur in addition to the
\emph{outer} turning points at $\omega_1t\!=\!\pi/2+\pi l$, when the
density profile is the broadest.  In Figs.~\ref{fig:big}\;\!(c) and
(f) we plot the half-width-at-half maximum (HWHM) $w(t)$ of the momentum distribution of the TG gas and compare it to the respective
result for an ideal Fermi gas, for which the narrowing occurs
only at the outer turning points. Away from the outer and inner turning points  the momentum distribution of both
TG and the ideal Fermi gas is
dominated by the hydrodynamic velocity (see below) 
and is 
homothetic to the Fermi-gas density profile
\cite{GangardtMinguzziExact,Rigol2005Fermionization}.

By comparing the dynamics of the  momentum widths 
shown in Figs.~\ref{fig:big}\;\!(c) and (f) at different temperatures, we observe the expected 
attenuation of the many-body bounce effect with increasing temperature as the
interactions become less important. 
For highly nondegenerate clouds, the
momentum distribution of  the TG gas converges towards that of the ideal Fermi gas, 
both being described by the Maxwell-Boltzmann distribution, so that one expects
the many-body bounce to be absent.
To characterise the dependence of the many-body bounce on both the
dimensionless initial temperature $\theta_0\!=\!k_B T_0/N\hbar\omega_0$ 
and the quench strength $\epsilon$, we introduce
the visibility parameter $K\!=\!I_{1}/I_{2}$ defined as the ratio between the
depths of local and global minima of the momentum width $w(t)$ at the inner and outer turning points, 
$\omega_{1}t=\pi$ and $\omega_{1}t=\pi/2$, respectively, as shown in Fig.~\ref{fig:big}\;\!(f).  In
terms of this parameter, the many-body bounce effect is the strongest for
$K\! \rightarrow \!1$, while $K \!=\! 0$ corresponds to its absence.

For a given
quench strength $\epsilon$ and total atom number $N$, we can define a dimensionless
crossover temperature $\theta_{\rm{cr}}\!=\!k_BT_0^{(\mathrm{cr})}/N\hbar\omega_0$ for
which the visibility parameter  $K$ attains a certain intermediate value. 
For convenience, we chose this value to be $K\!=\!0.366$ as for weak quenches
this corresponds to $w(t)$ being well approximated as
a sum of two sinusoidal harmonics of frequencies $\omega_B$ and
$2\omega_B$ and equal weights (see below).
In Fig.~\ref{Fig_Crossover_condition}, we plot the locations of
$\theta_{\rm{cr}}$ in the $\theta_0$--$\epsilon$ parameter space for different
$N$ and within the interval $-1\!<\!\epsilon \!<\!0$, corresponding to $\omega_1/\omega_0\!>\!1$ (i.e., tightening the trap).
As we see, $\theta_{\rm{cr}}$ barely depends on the atom number
for $N\gtrsim 8$, indicating that the thermodynamic limit is essentially reached and
that this figure can be regarded as a nonequilibrium crossover phase diagram of the
phenomenon of quantum many-body bounce in the breathing-mode oscillations of
the TG gas. We note that the phase diagram can be extended to the region $\epsilon\!>\!0$ (corresponding to $\omega_1/\omega_0\!<\!1$) by a transformation $\epsilon^{(>0)} = - \epsilon^{(<0)}/(1+\epsilon^{(<0)})$, which itself corresponds to inverting the value of $\omega_1/\omega_0\equiv r$ to $1/r$  \cite{phase-diagram}.

\begin{figure}[tbp]
\includegraphics[width=7.8cm]{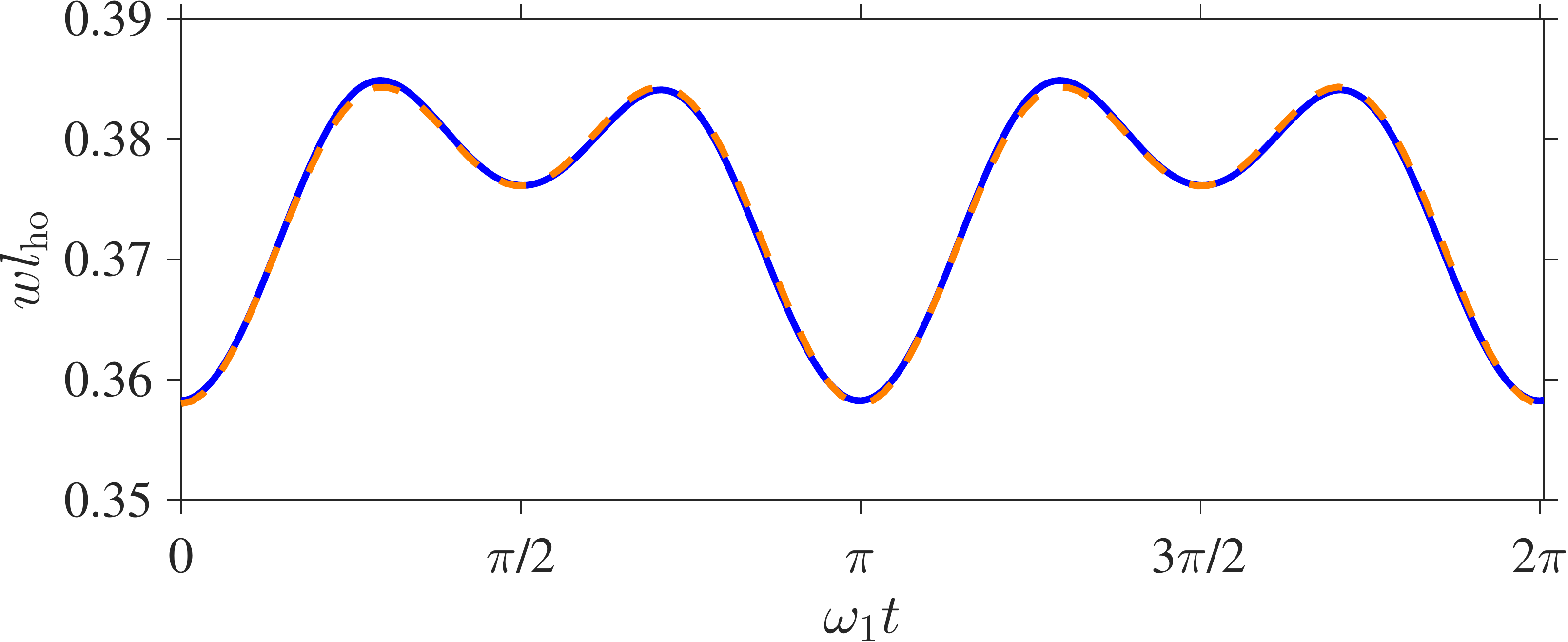}
\caption{(Color online) Dynamics of the width (HWHM) of the momentum distribution of the TG gas following a weak quench,
$\epsilon\!\simeq\!-0.0930 $ ($\omega_1\!=\!1.05\omega_0$), for $N\!=\!16$ and $\theta_0\!=\!0.01$.  
The solid (blue) line is the exact result, while the dashed (orange) line is based on a dual harmonic fit with frequencies $\omega_B$ and $2\omega_B$.
} 
\label{Fig_small_quench}
\end{figure}

In the thermodynamic limit, when 
the cloud size is much larger 
than the characteristic one-body correlation length, 
our exact results can be further understood by using the  local density 
approximation (LDA). 
 In the LDA, we can write   
$\rho_0(x,y) = \frac{1}{2\pi}\int dk \, e^{ik (x-y)} \,\bar{n} (k;\rho(X),T)$, 
where $X=(x+y)/2$ is the centre-of-mass coordinate 
and $\bar{n}(k;\rho,T)$ is the momentum distribution
of a uniform TG gas of density $\rho$ at temperature $T$, normalized to $\rho$. The quantity
$\bar{n}(k;\rho,T)$ can 
depend only on the dimensionless combinations, $k/\rho$ and
$k_BT/(\hbar^2\rho^2/m)$. Combining this with the LDA 
expression for $\rho_0(x,y)$ and the 
scaling solution (\ref{eq.ss}) leads to the following momentum distribution 
\begin{equation}
n(k,t)=\intop d x \, \bar{n}(k-mv(x,t)/\hbar;\rho(x,t),T(t)),
\label{eq:nkt-m}
\end{equation}
where $v(x,t) \!=\! x\dot\lambda(t) /\lambda(t)$ is the carrier hydrodynamic velocity field 
and $T(t) \!=\! T_0/\lambda(t)^2$ is the instantaneous temperature of the gas. 
Equation~(\ref{eq:nkt-m}) and the scaling solutions for $\rho(x,t)$, $v(x,t)$, and $T(t)$ are 
exactly the same as 
the ones that can be obtained directly from finite-temperature hydrodynamics of 1D Bose gases~\cite{Bouchoule_qbec:2016} (see~\cite{sup-Bounce}). This
equivalence stems from the existence of the scaling solutions for the
single-particle harmonic oscillator wavefunctions. We now present the hydrodynamics results and use them to gain additional insight into the 
physics of the quantum many-body bounce.

Neglecting the width of $\bar{n}$ in ~Eq.~(\ref{eq:nkt-m}) completely and approximating ~$\bar{n} (k-mv(x,t)/\hbar)$ ~by 
the delta-function ~~$\bar{n} (k-mv(x,t)/\hbar)\!\simeq \!\rho \;\!\delta (k-mv(x,t)/\hbar)$, yields the
following result for the  momentum distribution 
$n(k,t) \!= \!\frac{\hbar}{m |\dot{\lambda}|} \rho_0\!\left(  \frac{\hbar k }{m\dot{\lambda}} \right)$. 
The HWHM of this distribution [see Fig.~\ref{fig:big}\;\!(c)] 
vanishes both at the outer and inner turning points of the
breathing oscillations, where $\dot{\lambda}\!=\!0$. This corresponds to the
perfect many-body bounce, $K\!=\!1$. The vanishing of the width  
at the inner turning point occurs due to the increased pressure of the gas, acting as a
potential barrier, and represents a pure hydrodynamic manifestation of the
collective many-body bounce effect.

\begin{figure}[tbp] 
\includegraphics[width=8.3cm]{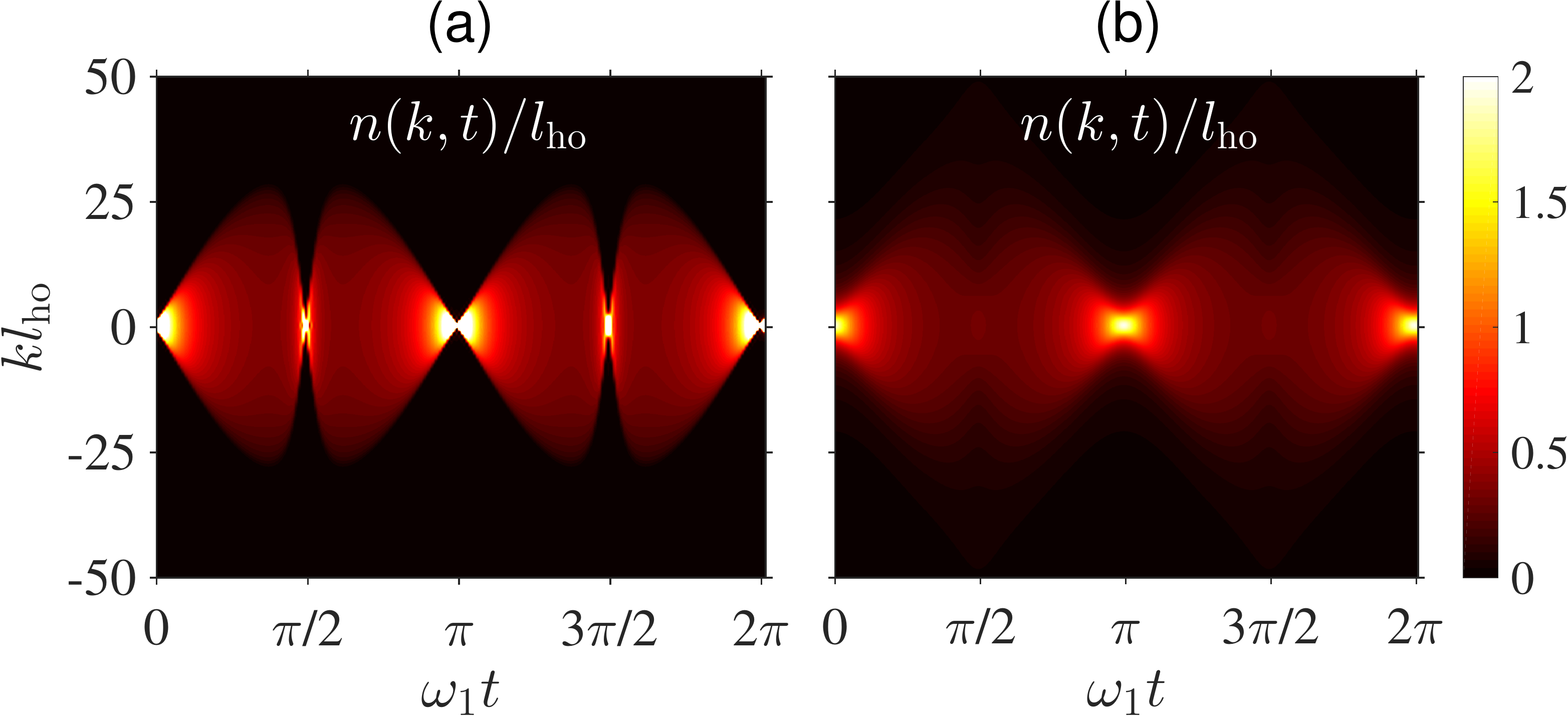}  
\caption{(Color online) Dynamics of the momentum distribution of the TG gas from the hydrodynamic approach 
and the Lorentzian approximation for $\bar{n}(k;\rho,T)$. The parameters are: $\epsilon\!=\!-0.9722 $ ($\omega_1\!=\!6\omega_0$), $\theta_0\!=\!0.01$ -- (a), and $\theta_0\!=\!0.5$  -- (b), with $\tilde{T}_0=\pi^2\theta_0$.
The dimensionless scale for the momentum axis, chosen here for direct comparison with Figs.~2\;\!(b) and (e) of the main text, 
corresponds to a choice of $k_BT_0/\hbar \omega_0$ that gives $N\!=\!16$.
}
\label{Fig:hydro}
\end{figure}

This picture becomes modified, however, if one takes into account thermal
broadening of the initial momentum distribution. 
Since the width of $\bar{n}(k;\rho,T)$ increases  with isentropic
compression, we expect that the width of $n(k,t)$ at the inner turning
points, where the gas is maximally compressed, is larger than at the
outer turning points, so that the  visibility parameter $K$ is reduced 
from its maximal value $K\!=\!1$.
 To see
this explicitly, we  model the momentum distribution of a 
uniform TG gas by a Lorentzian
$\bar{n}(k;\rho,T)\!=\!(2\rho l_{\phi}/\pi)/\left[1+(2l_{\phi}k)^{2}\right]$,
where $l_{\phi}\!=\!\hbar^{2}\rho /mk_{B}T $ is the  phase
coherence length \cite{Cazalilla:2004}. This expression, despite being valid
only for small momenta and low temperatures, $|k|\!\ll\! 1/l_\phi \!\ll\! \rho$, 
captures well the bulk of 
$\bar{n}$ and provides the dominant contribution
to the bulk of $n(k,t)$, Eq.~(\ref{eq:nkt-m}). 
Calculating $n(k,t)$ in this way \cite{sup-Bounce}  gives qualitatively good agreement with the exact results of 
Fig.~\ref{fig:big} and captures well the 
 temperature blurring of the many-body bounce (see Fig. \ref{Fig:hydro}).


In the weak quench regime, $|\epsilon|\!\ll\! 1$, the width $w(t)$ can be well described as a
sum of two harmonics with frequencies $\omega_B$ and $2\omega_B$, as shown in
Fig.~\ref{Fig_small_quench}. Due to the normalisation of the momentum
distribution, we expect that its peak at $k\!=\!0$ should oscillate 
out-of-phase with respect to $w(t)$.
Using the Lorentzian approximation for $\bar{n}(k; \rho, T)$, one can show (see \cite{sup-Bounce}) that the 
hydrodynamic result for $n(k,t)$ indeed leads to 
$n(0,t) \!\simeq\! a_0+a_1\cos(\omega_Bt)\!+\!a_2\cos(2\omega_Bt)$, with $a_1\!\simeq \!-4\epsilon Nl_\phi^{(0)}/3\pi^2$ and
$a_2\!\simeq \!128\epsilon^2 N l_\phi^{(0)} /105\pi^4\theta_0^2$, where
$l_\phi^{(0)}\!=\!\hbar^2\rho_0(0)/mk_B T_0$ is the phase coherence length in the trap centre.
 At low enough temperatures, the second harmonics, which
arises from the presence of the hydrodynamic velocity field $v(x,t)$,
dominates the oscillations. In this regime, the many-body bounce effect manifests itself as a
phenomenon of frequency doubling, observed recently in a weakly interacting
quasicondensate \cite{fang2014quench,Bouchoule_qbec:2016}.  Comparison of the magnitudes of
$a_1$ and $a_2$ allows us to also derive \cite{sup-Bounce} a simple scaling of $\theta_{\rm{cr}}\!\simeq \!0.3 \sqrt{|\epsilon|}$,
shown in the inset of Fig.~\ref{Fig_Crossover_condition}.

Note that the LDA result of Eq.~(\ref{eq:nkt-m}) also holds for an ideal
Fermi gas, provided that $\bar{n}\!=\!\bar{n}_\mathrm{F}$ is the
corresponding  momentum
distribution. 
The many-body bounce effect is, however,
absent in this case, even at arbitrarily low temperatures, because the
broadening of $\bar{n}_\mathrm{F}$ at the inner turning points due to the 
Pauli exclusion principle completely overwhelms the narrowing of the distribution of pure hydrodynamic
carrier velocities.
  We also point out 
  that the many-body
bounce in the TG gas cannot be revealed through the variance of the momentum
distribution,
$\langle (\Delta k)^2\rangle\!=\!\langle k^2 \rangle -\langle
k\rangle^2\!=\!\langle k^2\rangle$.
This quantity is proportional to the kinetic energy of the gas and is
therefore the same as in the ideal Fermi gas.  
Such a marked difference between the 
behavior of the variance and the HWHM 
is due to the fact the variance is dominated by the
contribution from the long $k^{-4}$ tails \cite{Minguzzi2002,Olshanii2003} of the TG gas, i.e., by 
momenta that are much larger than the HWHM.
For the ideal Fermi gas, on the other hand, both the variance and the HWHM are 
exhausted by the bulk of the momentum distribution.

In conclusion, we have shown that, in contrast to the dynamics of the
 \emph{in situ} density profile, the evolution of the momentum distribution of a harmonically
  quenched TG gas reveals a dramatic manifestation of interparticle
  interactions in the form a collective many-body bounce effect. 
  The many-body bounce should not be attributed to exclusively hard-core repulsion in the TG gas, but can also manifest itself as frequency doubling in the weakly interacting gases~\cite{fang2014quench,Bouchoule_qbec:2016}.
  This observation implies that collective effects in many-body systems 
can be better revealed via the dynamics of their momentum distribution, while not necessarily evident in the dynamics of the \emph{in situ} density profiles.

\begin{acknowledgments}
The authors acknowledge fruitful discussions with Y. Castin, E. Bogomolny, and O. Giraud.
I.\,B. acknowledges support by the Centre de Comp\'{e}tences Nanosciences \^{I}le-de-France.
K.\,V.\,K. acknowledges support by the Australian Research Council Discovery Project Grant
DP140101763.
\end{acknowledgments}



\begin{thebibliography}{44}%
\makeatletter
\providecommand \@ifxundefined [1]{%
 \@ifx{#1\undefined}
}%
\providecommand \@ifnum [1]{%
 \ifnum #1\expandafter \@firstoftwo
 \else \expandafter \@secondoftwo
 \fi
}%
\providecommand \@ifx [1]{%
 \ifx #1\expandafter \@firstoftwo
 \else \expandafter \@secondoftwo
 \fi
}%
\providecommand \natexlab [1]{#1}%
\providecommand \enquote  [1]{``#1''}%
\providecommand \bibnamefont  [1]{#1}%
\providecommand \bibfnamefont [1]{#1}%
\providecommand \citenamefont [1]{#1}%
\providecommand \href@noop [0]{\@secondoftwo}%
\providecommand \href [0]{\begingroup \@sanitize@url \@href}%
\providecommand \@href[1]{\@@startlink{#1}\@@href}%
\providecommand \@@href[1]{\endgroup#1\@@endlink}%
\providecommand \@sanitize@url [0]{\catcode `\\12\catcode `\$12\catcode
  `\&12\catcode `\#12\catcode `\^12\catcode `\_12\catcode `\%12\relax}%
\providecommand \@@startlink[1]{}%
\providecommand \@@endlink[0]{}%
\providecommand \url  [0]{\begingroup\@sanitize@url \@url }%
\providecommand \@url [1]{\endgroup\@href {#1}{\urlprefix }}%
\providecommand \urlprefix  [0]{URL }%
\providecommand \Eprint [0]{\href }%
\providecommand \doibase [0]{http://dx.doi.org/}%
\providecommand \selectlanguage [0]{\@gobble}%
\providecommand \bibinfo  [0]{\@secondoftwo}%
\providecommand \bibfield  [0]{\@secondoftwo}%
\providecommand \translation [1]{[#1]}%
\providecommand \BibitemOpen [0]{}%
\providecommand \bibitemStop [0]{}%
\providecommand \bibitemNoStop [0]{.\EOS\space}%
\providecommand \EOS [0]{\spacefactor3000\relax}%
\providecommand \BibitemShut  [1]{\csname bibitem#1\endcsname}%
\let\auto@bib@innerbib\@empty
\bibitem [{\citenamefont {Kinoshita}\ \emph {et~al.}(2006)\citenamefont
  {Kinoshita}, \citenamefont {Wenger},\ and\ \citenamefont
  {Weiss}}]{Kinoshita2006}%
  \BibitemOpen
  \bibfield  {author} {\bibinfo {author} {\bibfnamefont {T.}~\bibnamefont
  {Kinoshita}}, \bibinfo {author} {\bibfnamefont {T.}~\bibnamefont {Wenger}}, \
  and\ \bibinfo {author} {\bibfnamefont {D.~S.}\ \bibnamefont {Weiss}},\ }\href
  {http://dx.doi.org/10.1038/nature00968} {\bibfield  {journal} {\bibinfo
  {journal} {Nature}\ }\textbf {\bibinfo {volume} {440}},\ \bibinfo {pages}
  {900} (\bibinfo {year} {2006})}\BibitemShut {NoStop}%
\bibitem [{\citenamefont {Hofferberth}\ \emph {et~al.}(2007)\citenamefont
  {Hofferberth}, \citenamefont {Lesanovsky}, \citenamefont {Fischer},
  \citenamefont {Schumm},\ and\ \citenamefont
  {Schmiedmayer}}]{hofferberth2007}%
  \BibitemOpen
  \bibfield  {author} {\bibinfo {author} {\bibfnamefont {S.}~\bibnamefont
  {Hofferberth}}, \bibinfo {author} {\bibfnamefont {I.}~\bibnamefont
  {Lesanovsky}}, \bibinfo {author} {\bibfnamefont {B.}~\bibnamefont {Fischer}},
  \bibinfo {author} {\bibfnamefont {T.}~\bibnamefont {Schumm}}, \ and\ \bibinfo
  {author} {\bibfnamefont {J.}~\bibnamefont {Schmiedmayer}},\ }\href
  {http://dx.doi.org/10.1038/nature06149} {\bibfield  {journal} {\bibinfo
  {journal} {Nature}\ }\textbf {\bibinfo {volume} {449}},\ \bibinfo {pages}
  {324} (\bibinfo {year} {2007})}\BibitemShut {NoStop}%
\bibitem [{\citenamefont {Trotzky}\ \emph {et~al.}(2012)\citenamefont
  {Trotzky}, \citenamefont {Chen}, \citenamefont {Flesch}, \citenamefont
  {McCulloch}, \citenamefont {Schollw{\"o}ck}, \citenamefont {Eisert},\ and\
  \citenamefont {Bloch}}]{trotzky2012}%
  \BibitemOpen
  \bibfield  {author} {\bibinfo {author} {\bibfnamefont {S.}~\bibnamefont
  {Trotzky}}, \bibinfo {author} {\bibfnamefont {Y.-A.}\ \bibnamefont {Chen}},
  \bibinfo {author} {\bibfnamefont {A.}~\bibnamefont {Flesch}}, \bibinfo
  {author} {\bibfnamefont {I.~P.}\ \bibnamefont {McCulloch}}, \bibinfo {author}
  {\bibfnamefont {U.}~\bibnamefont {Schollw{\"o}ck}}, \bibinfo {author}
  {\bibfnamefont {J.}~\bibnamefont {Eisert}}, \ and\ \bibinfo {author}
  {\bibfnamefont {I.}~\bibnamefont {Bloch}},\ }\href
  {http://dx.doi.org/10.1038/nphys2232} {\bibfield  {journal} {\bibinfo
  {journal} {Nature Physics}\ }\textbf {\bibinfo {volume} {8}},\ \bibinfo
  {pages} {325} (\bibinfo {year} {2012})}\BibitemShut {NoStop}%
\bibitem [{\citenamefont {Gring}\ \emph {et~al.}(2012)\citenamefont {Gring},
  \citenamefont {Kuhnert}, \citenamefont {Langen}, \citenamefont {Kitagawa},
  \citenamefont {Rauer}, \citenamefont {Schreitl}, \citenamefont {Mazets},
  \citenamefont {Smith}, \citenamefont {Demler},\ and\ \citenamefont
  {Schmiedmayer}}]{gring2012}%
  \BibitemOpen
  \bibfield  {author} {\bibinfo {author} {\bibfnamefont {M.}~\bibnamefont
  {Gring}}, \bibinfo {author} {\bibfnamefont {M.}~\bibnamefont {Kuhnert}},
  \bibinfo {author} {\bibfnamefont {T.}~\bibnamefont {Langen}}, \bibinfo
  {author} {\bibfnamefont {T.}~\bibnamefont {Kitagawa}}, \bibinfo {author}
  {\bibfnamefont {B.}~\bibnamefont {Rauer}}, \bibinfo {author} {\bibfnamefont
  {M.}~\bibnamefont {Schreitl}}, \bibinfo {author} {\bibfnamefont
  {I.}~\bibnamefont {Mazets}}, \bibinfo {author} {\bibfnamefont {D.~A.}\
  \bibnamefont {Smith}}, \bibinfo {author} {\bibfnamefont {E.}~\bibnamefont
  {Demler}}, \ and\ \bibinfo {author} {\bibfnamefont {J.}~\bibnamefont
  {Schmiedmayer}},\ }\href {http://dx.doi.org/10.1126/science.1224953}
  {\bibfield  {journal} {\bibinfo  {journal} {Science}\ }\textbf {\bibinfo
  {volume} {337}},\ \bibinfo {pages} {1318} (\bibinfo {year}
  {2012})}\BibitemShut {NoStop}%
\bibitem [{\citenamefont {Fang}\ \emph {et~al.}(2014)\citenamefont {Fang},
  \citenamefont {Carleo}, \citenamefont {Johnson},\ and\ \citenamefont
  {Bouchoule}}]{fang2014quench}%
  \BibitemOpen
  \bibfield  {author} {\bibinfo {author} {\bibfnamefont {B.}~\bibnamefont
  {Fang}}, \bibinfo {author} {\bibfnamefont {G.}~\bibnamefont {Carleo}},
  \bibinfo {author} {\bibfnamefont {A.}~\bibnamefont {Johnson}}, \ and\
  \bibinfo {author} {\bibfnamefont {I.}~\bibnamefont {Bouchoule}},\ }\href
  {\doibase 10.1103/PhysRevLett.113.035301} {\bibfield  {journal} {\bibinfo
  {journal} {Phys. Rev. Lett.}\ }\textbf {\bibinfo {volume} {113}},\ \bibinfo
  {pages} {035301} (\bibinfo {year} {2014})}\BibitemShut {NoStop}%
\bibitem [{\citenamefont {Bloch}\ \emph {et~al.}(2008)\citenamefont {Bloch},
  \citenamefont {Dalibard},\ and\ \citenamefont
  {Zwerger}}]{Bloch-Dalibard-Zwerger}%
  \BibitemOpen
  \bibfield  {author} {\bibinfo {author} {\bibfnamefont {I.}~\bibnamefont
  {Bloch}}, \bibinfo {author} {\bibfnamefont {J.}~\bibnamefont {Dalibard}}, \
  and\ \bibinfo {author} {\bibfnamefont {W.}~\bibnamefont {Zwerger}},\ }\href
  {\doibase 10.1103/RevModPhys.80.885} {\bibfield  {journal} {\bibinfo
  {journal} {Rev. Mod. Phys.}\ }\textbf {\bibinfo {volume} {80}},\ \bibinfo
  {pages} {885} (\bibinfo {year} {2008})}\BibitemShut {NoStop}%
\bibitem [{\citenamefont {Cazalilla}\ and\ \citenamefont
  {Rigol}(2010)}]{CazalillaRigolNJP2010}%
  \BibitemOpen
  \bibfield  {author} {\bibinfo {author} {\bibfnamefont {M.~A.}\ \bibnamefont
  {Cazalilla}}\ and\ \bibinfo {author} {\bibfnamefont {M.}~\bibnamefont
  {Rigol}},\ }\href {http://stacks.iop.org/1367-2630/12/i=5/a=055006}
  {\bibfield  {journal} {\bibinfo  {journal} {New Journal of Physics}\ }\textbf
  {\bibinfo {volume} {12}},\ \bibinfo {pages} {055006} (\bibinfo {year}
  {2010})}\BibitemShut {NoStop}%
\bibitem [{\citenamefont {Polkovnikov}\ \emph {et~al.}(2011)\citenamefont
  {Polkovnikov}, \citenamefont {Sengupta}, \citenamefont {Silva},\ and\
  \citenamefont {Vengalattore}}]{polkovnikov2011}%
  \BibitemOpen
  \bibfield  {author} {\bibinfo {author} {\bibfnamefont {A.}~\bibnamefont
  {Polkovnikov}}, \bibinfo {author} {\bibfnamefont {K.}~\bibnamefont
  {Sengupta}}, \bibinfo {author} {\bibfnamefont {A.}~\bibnamefont {Silva}}, \
  and\ \bibinfo {author} {\bibfnamefont {M.}~\bibnamefont {Vengalattore}},\
  }\href {\doibase 10.1103/RevModPhys.83.863} {\bibfield  {journal} {\bibinfo
  {journal} {Rev. Mod. Phys.}\ }\textbf {\bibinfo {volume} {83}},\ \bibinfo
  {pages} {863} (\bibinfo {year} {2011})}\BibitemShut {NoStop}%
\bibitem [{\citenamefont {Cazalilla}\ \emph {et~al.}(2011)\citenamefont
  {Cazalilla}, \citenamefont {Citro}, \citenamefont {Giamarchi}, \citenamefont
  {Orignac},\ and\ \citenamefont {Rigol}}]{Cazalilla2011}%
  \BibitemOpen
  \bibfield  {author} {\bibinfo {author} {\bibfnamefont {M.~A.}\ \bibnamefont
  {Cazalilla}}, \bibinfo {author} {\bibfnamefont {R.}~\bibnamefont {Citro}},
  \bibinfo {author} {\bibfnamefont {T.}~\bibnamefont {Giamarchi}}, \bibinfo
  {author} {\bibfnamefont {E.}~\bibnamefont {Orignac}}, \ and\ \bibinfo
  {author} {\bibfnamefont {M.}~\bibnamefont {Rigol}},\ }\href {\doibase
  10.1103/RevModPhys.83.1405} {\bibfield  {journal} {\bibinfo  {journal} {Rev.
  Mod. Phys.}\ }\textbf {\bibinfo {volume} {83}},\ \bibinfo {pages} {1405}
  (\bibinfo {year} {2011})}\BibitemShut {NoStop}%
\bibitem [{lam()}]{lamacraft2012}%
  \BibitemOpen
  \href@noop {} {}\bibinfo {howpublished} {A. Lamacraft and J. Moore, in
  \emph{Ultracold bosonic and fermionic gases}, Vol. 5 (Contemporary Concepts
  in Condensed Matter Science), Eds. K. Levin, A. L. Fetter, and D. M.
  Stamper-Kurn (Elsevier, The Netherlands, 2012).}\BibitemShut {Stop}%
\bibitem [{\citenamefont {Pitaevskii}\ and\ \citenamefont
  {Stringari}()}]{Pitaevskii-Stringary-book}%
  \BibitemOpen
  \bibfield  {author} {\bibinfo {author} {\bibfnamefont {L.~P.}\ \bibnamefont
  {Pitaevskii}}\ and\ \bibinfo {author} {\bibfnamefont {S.}~\bibnamefont
  {Stringari}},\ }\href@noop {} {\emph {\bibinfo {title} {Bose-Einstein
  Condensation}}}\ (\bibinfo  {publisher} {Clarendon Press, Oxford,
  2003})\BibitemShut {NoStop}%
\bibitem [{\citenamefont {Stringari}(1996)}]{Stringari:1996}%
  \BibitemOpen
  \bibfield  {author} {\bibinfo {author} {\bibfnamefont {S.}~\bibnamefont
  {Stringari}},\ }\href {\doibase 10.1103/PhysRevLett.77.2360} {\bibfield
  {journal} {\bibinfo  {journal} {Phys. Rev. Lett.}\ }\textbf {\bibinfo
  {volume} {77}},\ \bibinfo {pages} {2360} (\bibinfo {year}
  {1996})}\BibitemShut {NoStop}%
\bibitem [{\citenamefont {Pitaevskii}\ and\ \citenamefont
  {Stringari}(1998)}]{Pitaevskii:1998}%
  \BibitemOpen
  \bibfield  {author} {\bibinfo {author} {\bibfnamefont {L.}~\bibnamefont
  {Pitaevskii}}\ and\ \bibinfo {author} {\bibfnamefont {S.}~\bibnamefont
  {Stringari}},\ }\href {\doibase 10.1103/PhysRevLett.81.4541} {\bibfield
  {journal} {\bibinfo  {journal} {Phys. Rev. Lett.}\ }\textbf {\bibinfo
  {volume} {81}},\ \bibinfo {pages} {4541} (\bibinfo {year}
  {1998})}\BibitemShut {NoStop}%
\bibitem [{\citenamefont {Jin}\ \emph {et~al.}(1996)\citenamefont {Jin},
  \citenamefont {Ensher}, \citenamefont {Matthews}, \citenamefont {Wieman},\
  and\ \citenamefont {Cornell}}]{JILA1996}%
  \BibitemOpen
  \bibfield  {author} {\bibinfo {author} {\bibfnamefont {D.~S.}\ \bibnamefont
  {Jin}}, \bibinfo {author} {\bibfnamefont {J.~R.}\ \bibnamefont {Ensher}},
  \bibinfo {author} {\bibfnamefont {M.~R.}\ \bibnamefont {Matthews}}, \bibinfo
  {author} {\bibfnamefont {C.~E.}\ \bibnamefont {Wieman}}, \ and\ \bibinfo
  {author} {\bibfnamefont {E.~A.}\ \bibnamefont {Cornell}},\ }\href {\doibase
  10.1103/PhysRevLett.77.420} {\bibfield  {journal} {\bibinfo  {journal} {Phys.
  Rev. Lett.}\ }\textbf {\bibinfo {volume} {77}},\ \bibinfo {pages} {420}
  (\bibinfo {year} {1996})}\BibitemShut {NoStop}%
\bibitem [{\citenamefont {Mewes}\ \emph {et~al.}(1996)\citenamefont {Mewes},
  \citenamefont {Andrews}, \citenamefont {van Druten}, \citenamefont {Kurn},
  \citenamefont {Durfee}, \citenamefont {Townsend},\ and\ \citenamefont
  {Ketterle}}]{MIT1996}%
  \BibitemOpen
  \bibfield  {author} {\bibinfo {author} {\bibfnamefont {M.-O.}\ \bibnamefont
  {Mewes}}, \bibinfo {author} {\bibfnamefont {M.~R.}\ \bibnamefont {Andrews}},
  \bibinfo {author} {\bibfnamefont {N.~J.}\ \bibnamefont {van Druten}},
  \bibinfo {author} {\bibfnamefont {D.~M.}\ \bibnamefont {Kurn}}, \bibinfo
  {author} {\bibfnamefont {D.~S.}\ \bibnamefont {Durfee}}, \bibinfo {author}
  {\bibfnamefont {C.~G.}\ \bibnamefont {Townsend}}, \ and\ \bibinfo {author}
  {\bibfnamefont {W.}~\bibnamefont {Ketterle}},\ }\href {\doibase
  10.1103/PhysRevLett.77.988} {\bibfield  {journal} {\bibinfo  {journal} {Phys.
  Rev. Lett.}\ }\textbf {\bibinfo {volume} {77}},\ \bibinfo {pages} {988}
  (\bibinfo {year} {1996})}\BibitemShut {NoStop}%
\bibitem [{\citenamefont {Jin}\ \emph {et~al.}(1997)\citenamefont {Jin},
  \citenamefont {Matthews}, \citenamefont {Ensher}, \citenamefont {Wieman},\
  and\ \citenamefont {Cornell}}]{JILA1997}%
  \BibitemOpen
  \bibfield  {author} {\bibinfo {author} {\bibfnamefont {D.~S.}\ \bibnamefont
  {Jin}}, \bibinfo {author} {\bibfnamefont {M.~R.}\ \bibnamefont {Matthews}},
  \bibinfo {author} {\bibfnamefont {J.~R.}\ \bibnamefont {Ensher}}, \bibinfo
  {author} {\bibfnamefont {C.~E.}\ \bibnamefont {Wieman}}, \ and\ \bibinfo
  {author} {\bibfnamefont {E.~A.}\ \bibnamefont {Cornell}},\ }\href {\doibase
  10.1103/PhysRevLett.78.764} {\bibfield  {journal} {\bibinfo  {journal} {Phys.
  Rev. Lett.}\ }\textbf {\bibinfo {volume} {78}},\ \bibinfo {pages} {764}
  (\bibinfo {year} {1997})}\BibitemShut {NoStop}%
\bibitem [{\citenamefont {Stamper-Kurn}\ \emph {et~al.}(1998)\citenamefont
  {Stamper-Kurn}, \citenamefont {Miesner}, \citenamefont {Inouye},
  \citenamefont {Andrews},\ and\ \citenamefont {Ketterle}}]{MIT1998}%
  \BibitemOpen
  \bibfield  {author} {\bibinfo {author} {\bibfnamefont {D.~M.}\ \bibnamefont
  {Stamper-Kurn}}, \bibinfo {author} {\bibfnamefont {H.-J.}\ \bibnamefont
  {Miesner}}, \bibinfo {author} {\bibfnamefont {S.}~\bibnamefont {Inouye}},
  \bibinfo {author} {\bibfnamefont {M.~R.}\ \bibnamefont {Andrews}}, \ and\
  \bibinfo {author} {\bibfnamefont {W.}~\bibnamefont {Ketterle}},\ }\href
  {\doibase 10.1103/PhysRevLett.81.500} {\bibfield  {journal} {\bibinfo
  {journal} {Phys. Rev. Lett.}\ }\textbf {\bibinfo {volume} {81}},\ \bibinfo
  {pages} {500} (\bibinfo {year} {1998})}\BibitemShut {NoStop}%
\bibitem [{\citenamefont {Moritz}\ \emph {et~al.}(2003)\citenamefont {Moritz},
  \citenamefont {St\"oferle}, \citenamefont {K\"ohl},\ and\ \citenamefont
  {Esslinger}}]{Moritz2003}%
  \BibitemOpen
  \bibfield  {author} {\bibinfo {author} {\bibfnamefont {H.}~\bibnamefont
  {Moritz}}, \bibinfo {author} {\bibfnamefont {T.}~\bibnamefont {St\"oferle}},
  \bibinfo {author} {\bibfnamefont {M.}~\bibnamefont {K\"ohl}}, \ and\ \bibinfo
  {author} {\bibfnamefont {T.}~\bibnamefont {Esslinger}},\ }\href {\doibase
  10.1103/PhysRevLett.91.250402} {\bibfield  {journal} {\bibinfo  {journal}
  {Phys. Rev. Lett.}\ }\textbf {\bibinfo {volume} {91}},\ \bibinfo {pages}
  {250402} (\bibinfo {year} {2003})}\BibitemShut {NoStop}%
\bibitem [{\citenamefont {Haller}\ \emph {et~al.}(2009)\citenamefont {Haller},
  \citenamefont {Gustavsson}, \citenamefont {Mark}, \citenamefont {Danzl},
  \citenamefont {Hart}, \citenamefont {Pupillo},\ and\ \citenamefont
  {N{\"a}gerl}}]{Naagerl2009}%
  \BibitemOpen
  \bibfield  {author} {\bibinfo {author} {\bibfnamefont {E.}~\bibnamefont
  {Haller}}, \bibinfo {author} {\bibfnamefont {M.}~\bibnamefont {Gustavsson}},
  \bibinfo {author} {\bibfnamefont {M.~J.}\ \bibnamefont {Mark}}, \bibinfo
  {author} {\bibfnamefont {J.~G.}\ \bibnamefont {Danzl}}, \bibinfo {author}
  {\bibfnamefont {R.}~\bibnamefont {Hart}}, \bibinfo {author} {\bibfnamefont
  {G.}~\bibnamefont {Pupillo}}, \ and\ \bibinfo {author} {\bibfnamefont
  {H.-C.}\ \bibnamefont {N{\"a}gerl}},\ }\href {\doibase
  10.1126/science.1175850} {\bibfield  {journal} {\bibinfo  {journal}
  {Science}\ }\textbf {\bibinfo {volume} {325}},\ \bibinfo {pages} {1224}
  (\bibinfo {year} {2009})}\BibitemShut {NoStop}%
\bibitem [{\citenamefont {Yuen}\ \emph {et~al.}(2015)\citenamefont {Yuen},
  \citenamefont {Barr}, \citenamefont {Cotter}, \citenamefont {Butler},\ and\
  \citenamefont {Hinds}}]{Hinds2016}%
  \BibitemOpen
  \bibfield  {author} {\bibinfo {author} {\bibfnamefont {B.}~\bibnamefont
  {Yuen}}, \bibinfo {author} {\bibfnamefont {I.}~\bibnamefont {Barr}}, \bibinfo
  {author} {\bibfnamefont {J.}~\bibnamefont {Cotter}}, \bibinfo {author}
  {\bibfnamefont {E.}~\bibnamefont {Butler}}, \ and\ \bibinfo {author}
  {\bibfnamefont {E.}~\bibnamefont {Hinds}},\ }\href
  {http://dx.doi.org/10.1088/1367-2630/17/9/093041} {\bibfield  {journal}
  {\bibinfo  {journal} {New Journal of Physics}\ }\textbf {\bibinfo {volume}
  {17}},\ \bibinfo {pages} {093041} (\bibinfo {year} {2015})}\BibitemShut
  {NoStop}%
\bibitem [{\citenamefont {Straatsma}\ \emph {et~al.}(2016)\citenamefont
  {Straatsma}, \citenamefont {Colussi}, \citenamefont {Davis}, \citenamefont
  {Lobser}, \citenamefont {Holland}, \citenamefont {Anderson}, \citenamefont
  {Lewandowski},\ and\ \citenamefont {Cornell}}]{JILA2016}%
  \BibitemOpen
  \bibfield  {author} {\bibinfo {author} {\bibfnamefont {C.~J.~E.}\
  \bibnamefont {Straatsma}}, \bibinfo {author} {\bibfnamefont {V.~E.}\
  \bibnamefont {Colussi}}, \bibinfo {author} {\bibfnamefont {M.~J.}\
  \bibnamefont {Davis}}, \bibinfo {author} {\bibfnamefont {D.~S.}\ \bibnamefont
  {Lobser}}, \bibinfo {author} {\bibfnamefont {M.~J.}\ \bibnamefont {Holland}},
  \bibinfo {author} {\bibfnamefont {D.~Z.}\ \bibnamefont {Anderson}}, \bibinfo
  {author} {\bibfnamefont {H.~J.}\ \bibnamefont {Lewandowski}}, \ and\ \bibinfo
  {author} {\bibfnamefont {E.~A.}\ \bibnamefont {Cornell}},\ }\href {\doibase
  10.1103/PhysRevA.94.043640} {\bibfield  {journal} {\bibinfo  {journal} {Phys.
  Rev. A}\ }\textbf {\bibinfo {volume} {94}},\ \bibinfo {pages} {043640}
  (\bibinfo {year} {2016})}\BibitemShut {NoStop}%
\bibitem [{\citenamefont {Minguzzi}\ \emph {et~al.}(2001)\citenamefont
  {Minguzzi}, \citenamefont {Vignolo}, \citenamefont {Chiofalo},\ and\
  \citenamefont {Tosi}}]{Minguzzi-hydro-2001}%
  \BibitemOpen
  \bibfield  {author} {\bibinfo {author} {\bibfnamefont {A.}~\bibnamefont
  {Minguzzi}}, \bibinfo {author} {\bibfnamefont {P.}~\bibnamefont {Vignolo}},
  \bibinfo {author} {\bibfnamefont {M.~L.}\ \bibnamefont {Chiofalo}}, \ and\
  \bibinfo {author} {\bibfnamefont {M.~P.}\ \bibnamefont {Tosi}},\ }\href
  {\doibase 10.1103/PhysRevA.64.033605} {\bibfield  {journal} {\bibinfo
  {journal} {Phys. Rev. A}\ }\textbf {\bibinfo {volume} {64}},\ \bibinfo
  {pages} {033605} (\bibinfo {year} {2001})}\BibitemShut {NoStop}%
\bibitem [{\citenamefont {Menotti}\ and\ \citenamefont
  {Stringari}(2002)}]{Menotti:2002}%
  \BibitemOpen
  \bibfield  {author} {\bibinfo {author} {\bibfnamefont {C.}~\bibnamefont
  {Menotti}}\ and\ \bibinfo {author} {\bibfnamefont {S.}~\bibnamefont
  {Stringari}},\ }\href {\doibase 10.1103/PhysRevA.66.043610} {\bibfield
  {journal} {\bibinfo  {journal} {Phys. Rev. A}\ }\textbf {\bibinfo {volume}
  {66}},\ \bibinfo {pages} {043610} (\bibinfo {year} {2002})}\BibitemShut
  {NoStop}%
\bibitem [{\citenamefont {Hu}\ \emph {et~al.}(2014)\citenamefont {Hu},
  \citenamefont {Xianlong},\ and\ \citenamefont {Liu}}]{Hu2014}%
  \BibitemOpen
  \bibfield  {author} {\bibinfo {author} {\bibfnamefont {H.}~\bibnamefont
  {Hu}}, \bibinfo {author} {\bibfnamefont {G.}~\bibnamefont {Xianlong}}, \ and\
  \bibinfo {author} {\bibfnamefont {X.-J.}\ \bibnamefont {Liu}},\ }\href
  {\doibase 10.1103/PhysRevA.90.013622} {\bibfield  {journal} {\bibinfo
  {journal} {Phys. Rev. A}\ }\textbf {\bibinfo {volume} {90}},\ \bibinfo
  {pages} {013622} (\bibinfo {year} {2014})}\BibitemShut {NoStop}%
\bibitem [{\citenamefont {Chen}\ \emph {et~al.}(2015)\citenamefont {Chen},
  \citenamefont {Li},\ and\ \citenamefont {Hu}}]{Hu2015}%
  \BibitemOpen
  \bibfield  {author} {\bibinfo {author} {\bibfnamefont {X.-L.}\ \bibnamefont
  {Chen}}, \bibinfo {author} {\bibfnamefont {Y.}~\bibnamefont {Li}}, \ and\
  \bibinfo {author} {\bibfnamefont {H.}~\bibnamefont {Hu}},\ }\href {\doibase
  10.1103/PhysRevA.91.063631} {\bibfield  {journal} {\bibinfo  {journal} {Phys.
  Rev. A}\ }\textbf {\bibinfo {volume} {91}},\ \bibinfo {pages} {063631}
  (\bibinfo {year} {2015})}\BibitemShut {NoStop}%
\bibitem [{\citenamefont {Choi}\ \emph {et~al.}(2015)\citenamefont {Choi},
  \citenamefont {Dunjko}, \citenamefont {Zhang},\ and\ \citenamefont
  {Olshanii}}]{Choi:2015}%
  \BibitemOpen
  \bibfield  {author} {\bibinfo {author} {\bibfnamefont {S.}~\bibnamefont
  {Choi}}, \bibinfo {author} {\bibfnamefont {V.}~\bibnamefont {Dunjko}},
  \bibinfo {author} {\bibfnamefont {Z.~D.}\ \bibnamefont {Zhang}}, \ and\
  \bibinfo {author} {\bibfnamefont {M.}~\bibnamefont {Olshanii}},\ }\href
  {\doibase 10.1103/PhysRevLett.115.115302} {\bibfield  {journal} {\bibinfo
  {journal} {Phys. Rev. Lett.}\ }\textbf {\bibinfo {volume} {115}},\ \bibinfo
  {pages} {115302} (\bibinfo {year} {2015})}\BibitemShut {NoStop}%
\bibitem [{\citenamefont {Gudyma}\ \emph {et~al.}(2015)\citenamefont {Gudyma},
  \citenamefont {Astrakharchik},\ and\ \citenamefont
  {Zvonarev}}]{Astrakharchik-Zvonarev-2015}%
  \BibitemOpen
  \bibfield  {author} {\bibinfo {author} {\bibfnamefont {A.~I.}\ \bibnamefont
  {Gudyma}}, \bibinfo {author} {\bibfnamefont {G.~E.}\ \bibnamefont
  {Astrakharchik}}, \ and\ \bibinfo {author} {\bibfnamefont {M.~B.}\
  \bibnamefont {Zvonarev}},\ }\href {\doibase 10.1103/PhysRevA.92.021601}
  {\bibfield  {journal} {\bibinfo  {journal} {Phys. Rev. A}\ }\textbf {\bibinfo
  {volume} {92}},\ \bibinfo {pages} {021601} (\bibinfo {year}
  {2015})}\BibitemShut {NoStop}%
\bibitem [{\citenamefont {De~Rosi}\ and\ \citenamefont
  {Stringari}(2015)}]{Stringari2015}%
  \BibitemOpen
  \bibfield  {author} {\bibinfo {author} {\bibfnamefont {G.}~\bibnamefont
  {De~Rosi}}\ and\ \bibinfo {author} {\bibfnamefont {S.}~\bibnamefont
  {Stringari}},\ }\href {\doibase 10.1103/PhysRevA.92.053617} {\bibfield
  {journal} {\bibinfo  {journal} {Phys. Rev. A}\ }\textbf {\bibinfo {volume}
  {92}},\ \bibinfo {pages} {053617} (\bibinfo {year} {2015})}\BibitemShut
  {NoStop}%
\bibitem [{\citenamefont {De~Rosi}\ and\ \citenamefont
  {Stringari}()}]{Stringari2016}%
  \BibitemOpen
  \bibfield  {author} {\bibinfo {author} {\bibfnamefont {G.}~\bibnamefont
  {De~Rosi}}\ and\ \bibinfo {author} {\bibfnamefont {S.}~\bibnamefont
  {Stringari}},\ }\href {https://arxiv.org/abs/1608.08417} {\bibinfo  {journal}
  {arXiv:1608.08417}\ }\BibitemShut {NoStop}%
\bibitem [{\citenamefont {Bouchoule}\ \emph {et~al.}(2016)\citenamefont
  {Bouchoule}, \citenamefont {Szigeti}, \citenamefont {Davis},\ and\
  \citenamefont {Kheruntsyan}}]{Bouchoule_qbec:2016}%
  \BibitemOpen
\bibfield  {journal} {  }\bibfield  {author} {\bibinfo {author} {\bibfnamefont
  {I.}~\bibnamefont {Bouchoule}}, \bibinfo {author} {\bibfnamefont {S.~S.}\
  \bibnamefont {Szigeti}}, \bibinfo {author} {\bibfnamefont {M.~J.}\
  \bibnamefont {Davis}}, \ and\ \bibinfo {author} {\bibfnamefont {K.~V.}\
  \bibnamefont {Kheruntsyan}},\ }\href {\doibase 10.1103/PhysRevA.94.051602}
  {\bibfield  {journal} {\bibinfo  {journal} {Phys. Rev. A}\ }\textbf {\bibinfo
  {volume} {94}},\ \bibinfo {pages} {051602} (\bibinfo {year}
  {2016})}\BibitemShut {NoStop}%
\bibitem [{\citenamefont {Girardeau}(1960)}]{Girardeau1960}%
  \BibitemOpen
  \bibfield  {author} {\bibinfo {author} {\bibfnamefont {M.}~\bibnamefont
  {Girardeau}},\ }\href {\doibase http://dx.doi.org/10.1063/1.1703687}
  {\bibfield  {journal} {\bibinfo  {journal} {Journal of {M}athematical
  {P}hysics}\ }\textbf {\bibinfo {volume} {1}},\ \bibinfo {pages} {516}
  (\bibinfo {year} {1960})}\BibitemShut {NoStop}%
\bibitem [{\citenamefont {Girardeau}(1965)}]{Girardeau_Bose_Fermi}%
  \BibitemOpen
  \bibfield  {author} {\bibinfo {author} {\bibfnamefont {M.~D.}\ \bibnamefont
  {Girardeau}},\ }\href {\doibase 10.1103/PhysRev.139.B500} {\bibfield
  {journal} {\bibinfo  {journal} {Phys. Rev.}\ }\textbf {\bibinfo {volume}
  {139}},\ \bibinfo {pages} {B500} (\bibinfo {year} {1965})}\BibitemShut
  {NoStop}%
\bibitem [{\citenamefont {Girardeau}\ and\ \citenamefont
  {Wright}(2000)}]{Girardeau2000}%
  \BibitemOpen
  \bibfield  {author} {\bibinfo {author} {\bibfnamefont {M.~D.}\ \bibnamefont
  {Girardeau}}\ and\ \bibinfo {author} {\bibfnamefont {E.~M.}\ \bibnamefont
  {Wright}},\ }\href {\doibase 10.1103/PhysRevLett.84.5239} {\bibfield
  {journal} {\bibinfo  {journal} {Phys. Rev. Lett.}\ }\textbf {\bibinfo
  {volume} {84}},\ \bibinfo {pages} {5239} (\bibinfo {year}
  {2000})}\BibitemShut {NoStop}%
\bibitem [{\citenamefont {Yukalov}\ and\ \citenamefont
  {Girardeau}(2005)}]{yukalov2005fermi}%
  \BibitemOpen
  \bibfield  {author} {\bibinfo {author} {\bibfnamefont {V.}~\bibnamefont
  {Yukalov}}\ and\ \bibinfo {author} {\bibfnamefont {M.}~\bibnamefont
  {Girardeau}},\ }\href {http://dx.doi.org/10.1002/lapl.200510011} {\bibfield
  {journal} {\bibinfo  {journal} {Laser Physics Letters}\ }\textbf {\bibinfo
  {volume} {2}},\ \bibinfo {pages} {375} (\bibinfo {year} {2005})}\BibitemShut
  {NoStop}%
\bibitem [{\citenamefont {Atas}\ \emph {et~al.}()\citenamefont {Atas},
  \citenamefont {Gangardt}, \citenamefont {Bouchoule},\ and\ \citenamefont
  {Kheruntsyan}}]{TonksMethods}%
  \BibitemOpen
  \bibfield  {author} {\bibinfo {author} {\bibfnamefont {Y.~Y.}\ \bibnamefont
  {Atas}}, \bibinfo {author} {\bibfnamefont {D.~M.}\ \bibnamefont {Gangardt}},
  \bibinfo {author} {\bibfnamefont {I.}~\bibnamefont {Bouchoule}}, \ and\
  \bibinfo {author} {\bibfnamefont {K.~V.}\ \bibnamefont {Kheruntsyan}},\
  }\href {https://arxiv.org/abs/1608.08720} {\bibinfo  {journal}
  {arXiv:1608.08720}\ }\BibitemShut {NoStop}%
\bibitem [{\citenamefont {Vignolo}\ and\ \citenamefont
  {Minguzzi}(2013)}]{vignolo2013universal}%
  \BibitemOpen
\bibfield  {journal} {  }\bibfield  {author} {\bibinfo {author} {\bibfnamefont
  {P.}~\bibnamefont {Vignolo}}\ and\ \bibinfo {author} {\bibfnamefont
  {A.}~\bibnamefont {Minguzzi}},\ }\href {\doibase
  10.1103/PhysRevLett.110.020403} {\bibfield  {journal} {\bibinfo  {journal}
  {Phys. Rev. Lett.}\ }\textbf {\bibinfo {volume} {110}},\ \bibinfo {pages}
  {020403} (\bibinfo {year} {2013})}\BibitemShut {NoStop}%
\bibitem [{\citenamefont {Perelomov}\ and\ \citenamefont
  {Zel'dovich}(1998)}]{PerelomovBook}%
  \BibitemOpen
  \bibfield  {author} {\bibinfo {author} {\bibfnamefont {A.}~\bibnamefont
  {Perelomov}}\ and\ \bibinfo {author} {\bibfnamefont {Y.}~\bibnamefont
  {Zel'dovich}},\ }\href@noop {} {\emph {\bibinfo {title} {Quantum
  {M}echanics}}}\ (\bibinfo  {publisher} {World Scientific, Singapore},\
  \bibinfo {year} {1998})\BibitemShut {NoStop}%
\bibitem [{\citenamefont {Minguzzi}\ and\ \citenamefont
  {Gangardt}(2005)}]{GangardtMinguzziExact}%
  \BibitemOpen
  \bibfield  {author} {\bibinfo {author} {\bibfnamefont {A.}~\bibnamefont
  {Minguzzi}}\ and\ \bibinfo {author} {\bibfnamefont {D.~M.}\ \bibnamefont
  {Gangardt}},\ }\href {\doibase 10.1103/PhysRevLett.94.240404} {\bibfield
  {journal} {\bibinfo  {journal} {Phys. Rev. Lett.}\ }\textbf {\bibinfo
  {volume} {94}},\ \bibinfo {pages} {240404} (\bibinfo {year}
  {2005})}\BibitemShut {NoStop}%
\bibitem [{sup()}]{sup-Bounce}%
  \BibitemOpen
  \href@noop {} {}\bibinfo {howpublished} {See the Supplemental Material at
  http://link.aps.org/ supplemental/XXX, which illustrates the breathing-mode
  dynamics for $N\!=\!2$ and outlines the details of the hydrodynamic
  solutions, calculation of the momentum distribution, and the weak quench
  expansion of $n(k\!=\!0,t)$.}\BibitemShut {Stop}%
\bibitem [{\citenamefont {Rigol}\ and\ \citenamefont
  {Muramatsu}(2005)}]{Rigol2005Fermionization}%
  \BibitemOpen
  \bibfield  {author} {\bibinfo {author} {\bibfnamefont {M.}~\bibnamefont
  {Rigol}}\ and\ \bibinfo {author} {\bibfnamefont {A.}~\bibnamefont
  {Muramatsu}},\ }\href {\doibase 10.1103/PhysRevLett.94.240403} {\bibfield
  {journal} {\bibinfo  {journal} {Phys. Rev. Lett.}\ }\textbf {\bibinfo
  {volume} {94}},\ \bibinfo {pages} {240403} (\bibinfo {year}
  {2005})}\BibitemShut {NoStop}%
\bibitem [{pha()}]{phase-diagram}%
  \BibitemOpen
  \href@noop {} {}\bibinfo {howpublished} {This observation follows from the
  fact that the breathing-mode dynamics, in appropriately scaled units, is
  exactly the same for $\epsilon\!>0$ as for $\epsilon \!<\!0$, provided that
  the quenches with negative and positive values of $\epsilon$ are related by
  $\epsilon^{(>0)} = - \epsilon^{(<0)}/(1+\epsilon^{(<0)})$ and that the
  dimensionless time axis $\tau=\omega_1t$ is shifted by $\pi/2$.}\BibitemShut
  {Stop}%
\bibitem [{\citenamefont {Cazalilla}(2004)}]{Cazalilla:2004}%
  \BibitemOpen
  \bibfield  {author} {\bibinfo {author} {\bibfnamefont {M.}~\bibnamefont
  {Cazalilla}},\ }\href {http://stacks.iop.org/0953-4075/37/i=7/a=051}
  {\bibfield  {journal} {\bibinfo  {journal} {Journal of Physics B: Atomic,
  Molecular and Optical Physics}\ }\textbf {\bibinfo {volume} {37}},\ \bibinfo
  {pages} {S1} (\bibinfo {year} {2004})}\BibitemShut {NoStop}%
\bibitem [{\citenamefont {Minguzzi}\ \emph {et~al.}(2002)\citenamefont
  {Minguzzi}, \citenamefont {Vignolo},\ and\ \citenamefont
  {Tosi}}]{Minguzzi2002}%
  \BibitemOpen
  \bibfield  {author} {\bibinfo {author} {\bibfnamefont {A.}~\bibnamefont
  {Minguzzi}}, \bibinfo {author} {\bibfnamefont {P.}~\bibnamefont {Vignolo}}, \
  and\ \bibinfo {author} {\bibfnamefont {M.~P.}\ \bibnamefont {Tosi}},\ }\href
  {\doibase 10.1016/S0375-9601(02)00042-7} {\bibfield  {journal} {\bibinfo
  {journal} {Phys. Lett. A}\ }\textbf {\bibinfo {volume} {294}},\ \bibinfo
  {pages} {222} (\bibinfo {year} {2002})}\BibitemShut {NoStop}%
\bibitem [{\citenamefont {Olshanii}\ and\ \citenamefont
  {Dunjko}(2003)}]{Olshanii2003}%
  \BibitemOpen
  \bibfield  {author} {\bibinfo {author} {\bibfnamefont {M.}~\bibnamefont
  {Olshanii}}\ and\ \bibinfo {author} {\bibfnamefont {V.}~\bibnamefont
  {Dunjko}},\ }\href {\doibase 10.1103/PhysRevLett.91.090401} {\bibfield
  {journal} {\bibinfo  {journal} {Phys. Rev. Lett.}\ }\textbf {\bibinfo
  {volume} {91}},\ \bibinfo {pages} {090401} (\bibinfo {year}
  {2003})}\BibitemShut {NoStop}%
\end{thebibliography}

%


\onecolumngrid\newpage



\section*{Supplementary material}

# Supplemental Material: Collective many-body bounce in the breathing-mode oscillations of a Tonks-Girardeau gas

Y. Y. Atas,[1] I. Bouchoule,[2] D. M. Gangardt,[3] and K. V. Kheruntsyan[1]

[1]*University of Queensland, School of Mathematics and Physics, Brisbane, Queensland 4072, Australia*
[2]*Laboratoire Charles Fabry, Institut d'Optique, CNRS, Univesité Paris Sud 11,*
*2 Avenue Augustin Fresnel, F-91127 Palaiseau Cedex, France*
[3]*School of Physics and Astronomy, University of Birmingham, Edgbaston, Birmingham, B15 2TT, UK*
(Dated: December 14, 2016)

## I. BREATHING-MODE OSCILLATIONS FOR $N=2$

To illustrate the breathing-mode oscillations in the simplest case of a two-body bounce ($N=2$), we plot in Fig. S1 the dynamics of the real-space density and momentum distributions in for $T_0 = 0$ and the same quench strength as in Fig. 1 of the main text. As we see, all qualitative features of the breathing-mode oscillations seen in Fig. 1 are preserved here, including the additional narrowing of the momentum distribution at the inner turning points of the breathing cycle, $\omega_1 t = \pi/2 + \pi l$ ($l = 1, 2, ...$). In Fig. S2, we plot a snapshot of the absolute-square of the two-particle wavefunction in momentum space, $|\psi(k_1, k_2; t)|^2$, at the inner turning point $\omega_1 t = \pi/2$ as a function of the relative and total momenta, for two hard-core bosons, and compare it with the respective result for two noninteracting fermions. From the essentially single-peaked structure of this probability distribution, centered at $k_1 - k_2 = k_1 + k_2 = 0$ for hard-core bosons, we can conclude that both particles slow down ($k_1 = k_2 = 0$) near the bottom of the trap and hence bounce off each other. In contrast, the probability distribution for two noninteracting fermions is double peaked at $k_1 - k_2 \neq 0$ and $k_1 + k_2 = 0$, implying that the particles simply "pass" through each other without slowing down, with equal but opposite momenta ($k_1 = -k_2 \neq 0$), just like the classical ideal gas particles would do.

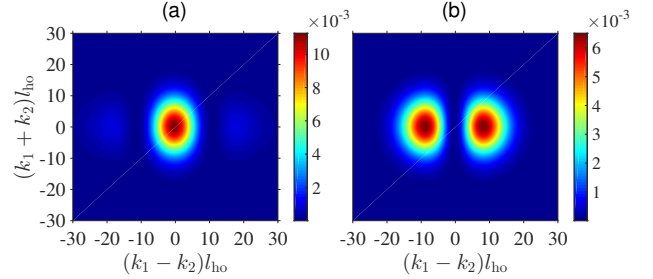

FIG. S2. (Color online) Snapshots of the absolute square of the two-particle wavefunction in momentum space, $|\psi(k_1, k_2; t = \pi/2\omega_1)|^2$, at the inner turning point of the breathing cycle ($\omega_1 t = \pi/2$), for (a) two hard-core bosons and (b) two noninteracting fermions, for $T_0 = 0$ and $\epsilon \simeq -0.9722$ ($\omega_1 = 6\omega_0$).

## II. HYDRODYNAMIC APPROACH AND WEAK QUENCH EXPANSION OF $n(0,t)$

The hydrodynamic equations describing our system are given by [1]

$$\partial_t \rho + \partial_x(\rho v) = 0, \quad (S1)$$

$$\partial_t v + v \partial_x v = -\frac{1}{m}\partial_x V(x,t) - \frac{1}{m\rho}\partial_x P, \quad (S2)$$

$$\partial_t s + v \partial_x s = 0, \quad (S3)$$

and rely on the assumption that the TG gas can be divided into small locally uniform slices, each of which maintains local thermal equilibrium and undergoes isentropic expansion and compression cycles. Here, $\rho(x,t)$ is the local 1D density of the slice at position $x$, $v(x,t)$ is the respective hydrodynamic velocity, $s(x,t)$ is the entropy per particle, $V(x,t) = \frac{1}{2}m\omega(t)^2 x^2$ is the external trapping potential as before, and $P(x,t)$ is the local thermodynamic pressure.

The hydrodynamic equations in a given trap potential depend only of the thermodynamic equation of state for the pressure. As the TG and ideal Fermi gases share the same equation of state, the solutions to the hydrodynamic equations for the TG gas are exactly the same as for the equivalent ideal Fermi gas. Using dimensional arguments as in Ref. [1] (applicable to either bosonic or fermionic ideal gases at arbitrary temperatures), one can show that Eqs. (S1)-(S3) are satisfied by the scaling solutions of the form

$$\rho(x,t) = \rho_0(x/\lambda(t))/\lambda(t), \quad v(x,t) = x\dot{\lambda}(t)/\lambda(t), \quad (S4)$$

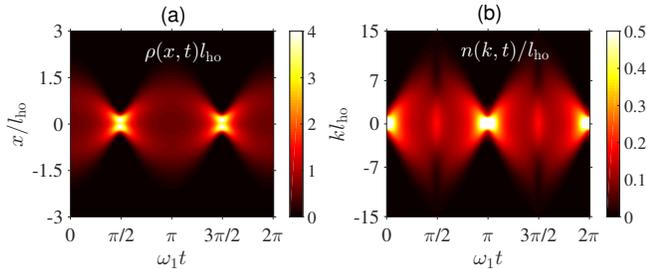

FIG. S1. (Color online) Breathing-mode dynamics of (a) the real-space density and (b) the momentum distribution of the TG gas for $N = 2$, $T_0 = 0$, and $\epsilon \simeq -0.9722$ ($\omega_1 = 6\omega_0$).

$$T(t) = T_0/\lambda(t)^2, \tag{S5}$$

which are the same ones as in the main text [see Eq. (1) and the text after Eq. (5)], except that we collate them here again for convenience.

To describe the evolution of the bulk of the momentum distribution $n(k,t)$ of the TG gas at temperatures well below the temperature of quantum degeneracy, we use Eq. (5) of the main text, together with the Lorentzian approximation for $\bar{n}(k; \rho(x,t), T(t))$. In addition, we need the density profile $\rho(x,t)$, which is obtained with the help of Eq. (1) of the main text from the initial profile $\rho_0(x)$; the latter can be approximated by the Thomas-Fermi (TF) semicircle $\rho_0(x) = \rho_0(0)\sqrt{1-x^2/R_{\text{TF}}^2}$ for $k_B T_0 \ll \hbar^2 \rho_0(0)^2/m$, where $R_{\text{TF}} = 2N/\pi\rho_0(0) = \sqrt{2N}l_{\text{ho}}$ is the TF size of the cloud. Implementing this leads to the following hydrodynamics result for $n(k,t)$, in a dimensionless form:

$$\frac{n(k,t)}{Nl_\phi^{(0)}} = \frac{4\tilde{\lambda}}{\pi^2}\int_{-1}^{1}du\,\frac{(1-u^2)}{1+4\tilde{\lambda}^2(1-u^2)\left(\tilde{k}-2\frac{\omega_1}{\omega_0}\dot{\tilde{\lambda}}u/\tilde{T}_0\right)^2}. \tag{S6}$$

Here, $l_\phi^{(0)} = \hbar^2\rho_0(0)/mk_B T_0$ is the initial phase correlation length in the trap centre, whereas $\tilde{T}_0 = T_0/T_d$ is the dimensionless temperature, with $T_d = \hbar^2\rho_0(0)^2/(2mk_B)$ being the initial temperature of quantum degeneracy of a uniform 1D gas at density $\rho_0(0)$. In addition, $\tilde{\lambda}(\tau) \equiv \lambda(\tau/\omega_1)$, where $\tau = \omega_1 t$ is the dimensionless time, so that $\dot{\tilde{\lambda}} = d\tilde{\lambda}/d\tau = \dot{\lambda}(t)/\omega_1$ and $\tilde{\lambda}(\tau) = [1+\epsilon\sin^2(\tau)]^{1/2}$. The normalisation of the density profile to the total atom number $N$ results in the following useful relationship (used in the comparison between the hydrodynamic results and the exact results for fixed $N$) between the harmonic oscillator length $l_{\text{ho}}$ and the phase correlation length $l_\phi^{(0)}$: $l_{\text{ho}} = \frac{1}{\pi}\sqrt{\frac{N}{2}}\,\tau_0 l_\phi^{(0)}$, with $\tilde{T}_0 = \pi^2\theta_0$ and $\rho_0(0) = \sqrt{2N}/\pi l_{\text{ho}}$. As we see, apart from the normalisation factor, equation (S6)—for a given ratio of $\omega_1/\omega_0$ (or, equivalently, for a given quench strength $\epsilon$)—depends only on a single dimensionless parameter $\tilde{T}_0$.

Examples of evolution of the momentum distribution of a finite-temperature TG gas, calculated using Eq. (S6), are shown in Fig. 4 of the main text, for $\epsilon \simeq -0.9722$ ($\omega_1 = 6\omega_0$) and two different temperatures. As we see, the Lorentzian approximation for $\bar{n}(k;\rho,T)$ leads to a good qualitative agreement of the hydrodynamic result for $n(k,t)$ with the exact results of Fig. 1 of the main text

For a weak quench, $|\epsilon| \ll 1$, we can expand Eq. (2) of the main text to obtain,

$$\lambda(t) \simeq 1 + \frac{\epsilon}{4}[1-\cos(\omega_B t)], \tag{S7}$$

$$\dot{\lambda}^2 \simeq \frac{\epsilon^2}{8}\omega_1^2[1-\cos(2\omega_B t)]. \tag{S8}$$

This in turn allows one to identify the two harmonics at the frequencies $\omega_B = 2\omega_1$ and $2\omega_B$ in the expansion of the peak momentum distribution:

$$n(0,t) \simeq \frac{2\hbar^2\lambda}{\pi mk_B T_0}\int dy\rho_0^2(y)\left[1-\left(\frac{2\hbar\dot{\lambda}\lambda y\rho_0(y)}{k_B T_0}\right)^2\right]$$
$$\simeq a_0 + a_1\cos(\omega_B t) + a_2\cos(2\omega_B t). \tag{S9}$$

In the lowest order in $\epsilon$ the expansion coefficients are given by

$$a_1 \simeq -\epsilon\frac{a_0}{4} \simeq -\epsilon\frac{\hbar^2}{2\pi mk_B T_0}\int dy\rho_0^2(y), \tag{S10}$$

$$a_2 \simeq \epsilon^2\frac{\hbar^2}{\pi mk_B T_0}\left(\frac{\hbar\omega_1}{k_B T_0}\right)^2\int dy\, y^2\rho_0^4(y). \tag{S11}$$

We now use the TF semicircle density profile to calculate the integrals in Eqs. (S10) and (S11) explicitely. This gives

$$a_1 \simeq -\frac{4\epsilon}{3\pi^2}Nl_\phi^{(0)},\; a_2 \simeq \frac{128\epsilon^2}{105\pi^4}\left(\frac{N\hbar\omega_1}{k_B T_0}\right)^2 Nl_\phi^{(0)}. \tag{S12}$$

Equating the magnitudes of $a_1$ and $a_2$ to each other, which is equivalent to our definition of the crossover temperature $\theta_{\text{cr}} = k_B T_0^{(\text{cr})}/N\hbar\omega_0$, leads to the following simple scaling $\theta_{\text{cr}} \simeq \sqrt{32|\epsilon|/35\pi^2} \simeq 0.3\sqrt{|\epsilon|}$ (shown in the inset of Fig. 3 of the main text) of the crossover conditions for observing the many-body bounce effect for $|\epsilon| \ll 1$.

---

[1] I. Bouchoule, S. S. Szigeti, M. J. Davis, and K. V. Kheruntsyan, Phys. Rev. A **94**, 051602 (2016).



\end{document}